\documentclass{llncs}
\usepackage{graphicx}
\usepackage{epsfig}
\usepackage{amsmath}
\usepackage{rotating}
\usepackage{adjustbox}
\usepackage{blindtext}
\usepackage{placeins}
\begin{document}

\title{Machine Learning in Astronomy: A Case Study in Quasar-Star Classification}
\titlerunning{Machine Learning for Quasar-Star Classification}  
%
\author{Mohammed Viquar\inst{1} \and Suryoday Basak\inst{1}\and 
Ariruna Dasgupta \inst{2}\and Surbhi Agrawal\inst{1} \and Snehanshu Saha \inst{1}}
\authorrunning{Mohammed Viquar et al.} 
%
\tocauthor{Mohammed Viquar, Suryoday Basak, Ariruna Dasgupta, Surbhi Agrawal, Snehanshu Saha}
\institute{Department of Computer Science and Engineering, and Center for Applied Mathematical Modeling and Simulation (CAMMS), PESIT South Campus, Bangalore, India,\\
\and
Dept. of Computer Science, University of Calcutta, India.\\
\email{viquar27x4@gmail.com, suryodaybasak@pes.edu, dasguptaariruna@gmail.com,\\surbhiagrawal@pes.edu, snehanshusaha@pes.edu}\\
WWW Homepage: \texttt{http://astrirg.org}}

\maketitle              

\begin{abstract}
We present the results of various automated classification methods, based on machine learning (ML), of objects from data releases 6 and 7 (DR6 and DR7) of the Sloan Digital Sky Survey (SDSS), primarily distinguishing stars from quasars. We provide a careful scrutiny of approaches available in the literature and have highlighted the pitfalls in those approaches based on the nature of data used for the study. The aim is to investigate the appropriateness of the application of certain ML methods. The manuscript argues convincingly in favor of the efficacy of asymmetric AdaBoost to classify photometric data. The paper presents a critical review of existing study and puts forward an application of asymmetric AdaBoost, as an offspring of that exercise.
\end{abstract}
%
\section{Introduction} \label{sec:intro}
A quasar is a \textit{quasi-stellar radio source}, which was first discovered in 1960. They emit electromagnetic radiation in the frequency bands corresponding to radio waves, visible, ultraviolet, infrared, X-rays and gamma rays. They are many light-years away from the Earth and the radiation from a quasar could take billions of years to reach us and may carry signatures of the early stages of the universe. This information gathering exercise and subsequent physical analysis of quasars pose strong motivation for the current study. It is difficult for astronomers to study quasars by relying on telescopic observations with template manual matching alone since quasars are difficult to distinguish from stars due to their great distance from Earth. Hence, in this paper, we present methods which can be scaled up to semi-automated or automated techniques to distinguish quasars from stars.

\textit{Machine learning} (ML) \cite{Basak :c11} is a sub-field of computer science which relies on statistical methods for predictive analysis. \textit{Supervised} ML algorithms rely on a representative sample of data to make predictions of class-belongingness for new or incoming data. The methods we elucidate in this paper use supervised machine learning approaches with proper bias handling in the data. The ML algorithms that we have tried are support vector machines (SVM), SVM and $K$ nearest neighbor hybrid (SVM-KNN), AdaBoost, and asymmetric AdaBoost. Of these four methods, SVM and SVM-KNN have been previously tried for quasar-star classification, but we improve upon the performance (and the justification for using them) by introducing methods of bias-handling. To the best of our knowledge, AdaBoost and asymmetric AdaBoost have not been previously tried to solve this problem. To contrast the effects of bias that arise due to the imbalance in the data, we have performed the experiments on naturally imbalanced as well as artificially balanced data sets.

The outcome of this research is two-fold. The first, to assert appropriate models for the separation of stars and quasars; and the second, to provide a solid reasoning for selecting these models, and consequently establishing a set of best practices for data scientific research in astronomy.
\section{Literature Survey} \label{sec:litsurvey}
Support vector machine (SVM) is one of the most widely used and powerful ML methods.  The authors in \cite{Gao:c5} attempted to solve the quasars-stars classification problem by using SVM to classify the star and quasar samples that are present in the Sloan Digital Sky Survey (SDSS) database. \cite{Elting:c6} used SVM for classifying stars, galaxies, and quasars. Both of them use nonlinear radial basis function (RBF or Gaussian) kernel. Although the accuracies reported were high, a justification of selection of the RBF kernel was not forthcoming. The authors in the present manuscript have performed a linear separability test on the data set, discussed in Section \ref{sec:method}, which clearly shows that the data is mostly linearly separable and hence, a linear SVM can be used. \cite{Peng et al.(2013):c7} used an SVM-KNN method which is a combination of SVM and KNN. SVM-KNN improves the performance of SVM by using KNN to better classify the samples which occur near the boundary (hyperplane) constructed by the SVM learner. In other works, decision tree classifiers are also used for star-galaxy separation \cite{Eduardo et al. (2017):c13}.

If data are linearly separable, then SVM may be implemented using a linear kernel. The absence of linear separability may justify SVM implementation in conjunction with the RBF kernel. In \cite{Elting:c6}, \cite{Gao:c5} and \cite{Peng et al.(2013):c7}, such an exploration is not reported. Moreover, the class dominance was ignored by \cite{Elting:c6}, \cite{Gao:c5} and \cite{Peng et al.(2013):c7}. Class dominance must be considered; otherwise, the accuracy of classification obtained will be biased by the dominant class and it will always be numerically very high. We have performed \textit{artificial balancing} of data to counter the effects of class bias; the process of artificial balancing has been elaborated in \ref{subsec:methods_art_balancing}. In addition to using previously tried ML models with improvements on bias-handling, in this paper, we explore asymmetric AdaBoost, which is a method designed to handle imbalanced datasets.
\section{Data Acquisition}\label{sec:dataacq}
The Sloan Digital Sky Survey (SDSS) is the most extensive redshift survey of the Universe, whose data collection began in 1998. Data release (here on, just \textit{DR}) 6 \cite{Adelman:c1}, comprises of the complete imaging over the northern Galactic cap. As a part of this survey, about 287 million objects are registered, over 9583 deg$^2$. More than 1.27 million spectra are available from this survey in the $u$, $g$, $r$, $i$ and $z$ bands. DR7 \cite{Abazajian:c2}, released in 2009, covers 11,663 deg$^2$ of the sky. The DR7 was the end of the SDSS-II phase. This catalog contains the same five bands of data as in the DR6, but of 357 million distinct objects. All of the data that is released by SDSS is made available over the Internet \cite{sdss}. The SkyServer provides interfaces for querying and obtaining data as per a user's needs. Using the available interfaces, spectral data, as well as images, can be obtained \cite{cosmos}. The data are available for non-commercial use only, without written permission. From this data, we make use of the classes of quasars and stars and extend the work done by \cite{Elting:c6}, \cite{Gao:c5} and \cite{Peng et al.(2013):c7}.
\section{Method}\label{sec:method}
\subsection{Artificial Balancing of Data \label{subsec:methods_art_balancing}}
\textit{Artificial balancing} of data needs to be performed such that the classes present in the dataset used for training a model don not present a bias to the learning algorithm. The ratio of the number of quasars to the number of stars is 7:1, and hence, either class is not equally represented to the classifier. In quasar-star classification, the stars' class dominates the quasars' class. This causes an increase in the influence of the stars' class on the learning algorithm and results in a higher accuracy of classification. In artificial balancing, an equal number of samples from both the classes are taken for training the classifier. This eliminates the class bias and the data imbalance. 

Without artificial balancing, the dataset used for analysis uses a larger number of samples belonging to the stars' class as compared to the number of samples in the quasars' class. The samples that are classified as belonging to the stars' class are more when compared to the number of samples classified as belonging to the quasars' class as the voting for the dominating class increases with imbalance and results in a higher accuracy of classification. Hence, the voting for the stars' class was found to be 99.41\% which is higher than the voting of quasars, which is 98.19\%, by \cite{Peng et al.(2013):c7}. The accuracy claimed is doubtful as data imbalance and class bias is prevalent.
%
%
\subsection{Separability Test}\label{subsec:separability_test}

\begin{figure}[ht]
  \begin{adjustbox}{addcode={\begin{minipage}{\width}}{\caption{%
      Convex hulls across every pair of features show that the two classes can be approximately wrapped into two separate, non-overlapping polygons when considering redshift as a feature. The data points belonging to the class of quasars are plotted in red, and those belonging to the class of stars are plotted in blue.
      }\end{minipage}},center}
      \includegraphics[scale=.48]{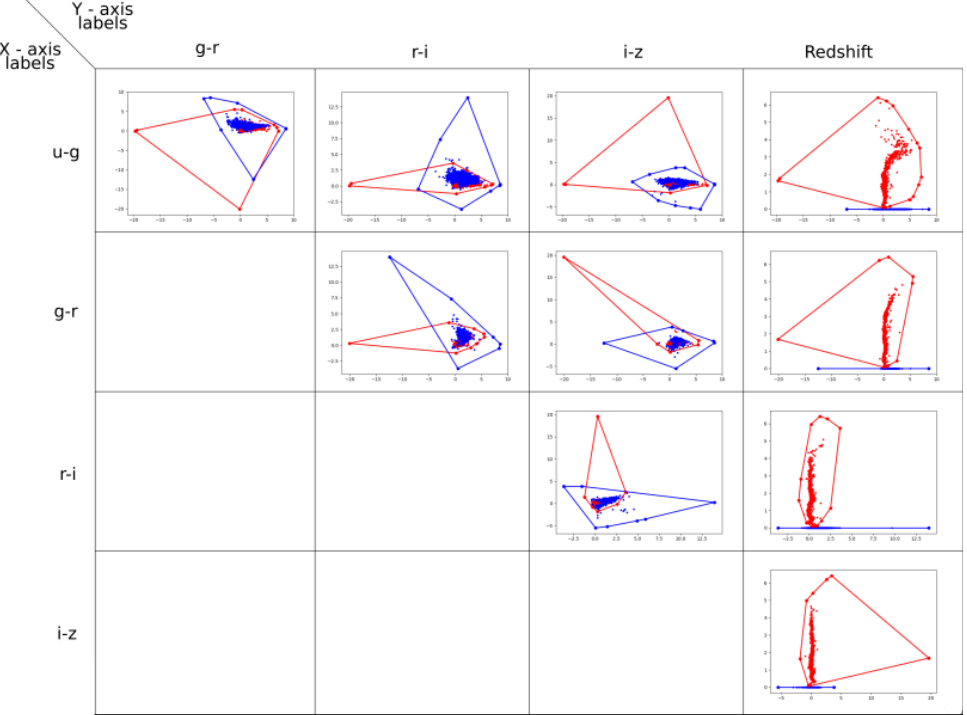}%
  \end{adjustbox}
\end{figure}

A \textit{separability test} is used to determine the nature of the separability of data. In particular, if the data are not linearly separable, certain classifiers may not work well or may not be appropriate.

\FloatBarrier

The \textit{convex hulls} of different classes in the dataset provides us with an indication of separability: the convex hull of a given set of points is the smallest \textit{n}-dimensional polygon which can adequately envelope all the points in the respective set. In general, if the convex hull of at least any two classes of any data set intersects or overlaps, then it may be concluded that the classes in the data are not linearly separable.

In the existing literature on quasars-stars classification, there a strong justification is not provided for the use of an RBF kernel. However, in the Figure 1, is observed that the majority of the data belonging to the class of stars are not present within the convex hull of the class of quasars. Thus, the two classes in the dataset are mostly linearly separable and SVM can be used here. Since the data exhibits linear separability, an RBF kernel need not be used.

\subsection{Support Vector Machine}\label{subsec:svm}
An SVM classifier requires the data to be separable so that it is possible to yield a hyperplane separating both the classes. Consider a set of $n$ samples from the data set and two classes $C_1$ and $C_2$ corresponding to quasars and stars, or vice versa. Let $x$ be the input matrix with $n$ rows corresponding to the $n$ data points and an array $y$ with $n$ elements, where the $j^{th}$ element of $y$ is the class label of the $j^{th}$ row in $x$. Out of the set of $n$ points, a pair of points, created by taking one from either class, is used to create a support vector $S$. Each point is then added to the support vector $S$. The position of samples from both the classes are determined in a 5-dimensional support vector (the five dimensions being $u-g$, $g-r$, $r-i$, $i-z$, and $Z$); any points which are geometrically present on the wrong side of the hyperplane by virtue of their class belongingness are added to a vector $V$ such that $S = S \cup V$. If any coefficients are negative due to the addition of $V$ to $S$, then such points are pruned.
\subsection{SVM-KNN \label{subsec:svm_knn}}
The K-nearest neighbor (KNN) classifier is a simple method for algorithmic classification, based on geometric similarity of the $K$ closest training samples in the feature space. When a previously unobserved sample is fed for classification to the $K-NN$ classifier, it searches the feature space for the $K$ samples which are closest to the test sample. The $K$ closest samples may belong to different classes; the learning algorithm selects the class to which the majority of the $K$ nearest samples belong and determines it to be the class to which the test sample belongs. Here, the parameter $K$ needs to be fed as an input and often depends on the data being explored. However, in practice, a value of $K$ between 7 and 11 works well \cite{Hassanat et al.}.
\subsection{AdaBoost}\label{subsec:adaboost}
Adaptive Boost or \textit{AdaBoost} \cite{Freund:c8} is a general ensemble learning approach that makes use of the results of multiple weak learners to make a strong prediction. AdaBoost works in multiple rounds by incrementally training weak learners, where each successive weak learner tries to classify the misclassified samples of the previous learner, with increased weights on the misclassified samples. AdaBoost can be used on any learning algorithm but the most popular learners for AdaBoost are short \textit{decision trees} or \textit{decision stumps}\cite{Basak :c11}. In the current study, the weak learners over which AdaBoost was used are decision trees with one level.
\subsection{Asymmetric AdaBoost: Handling the Data Imbalance Problem Mathematically\label{subsec:assymadab}}
The asymmetric AdaBoost algorithm \cite{Alba-Castro(2012):c9} aims to incorporate initial costs of misclassification in order to make the AdaBoost algorithm more sensitive to biases.

Consider a set of $n$ training samples $(\mathbf{x}_i, y_i; i = 1, 2, \dots, n)$ where $\mathbf{x}_i$ and $y_i$ are the feature vector, and class label of the $i^{th}$ sample respectively. Without loss of generality, it can be assumed that, the first $m$ examples have class label $y_i = 1; i = 1, 2, \dots, m$, and the remaining $n-m$ examples have class label $y_i = -1; i = m+1, m+2, \dots, n$, corresponding to the classes of quasars and stars respectively. Here, $m = 74,463$ and $n-m = 430,827$. Let us define a weight distribution $D_t(i); t = 1, 2, \dots; i = 1, 2, \dots , n$ over the whole training set where the index $t$ denotes the $t^{th}$ iteration of the AdaBoost algorithm, and the total number of iterations is equal to 1,000. The weak learner selects the best classifier according to the weight distribution. In regular AdaBoost, the initial weights are usually assigned as $D_t(i) = 1/n$, $\forall i$. After each iteration, the weight distribution is modified in such a way that misclassified samples get a higher penalty than the correctly classified samples: this is similar to regular AdaBoost. However, an asymmetric behavior is observed in AdaBoost: while updating weights in successive iterations, it treats the misclassification of positive samples and negative samples equally. But there may be situations where misclassification of a positive sample may be more expensive than that of a negative sample, which introduces an asymmetry to the problem. Asymmetry can also be introduced when the number of samples belonging to one class dominates over that of the other. The classification power of regular AdaBoost diminishes as such asymmetry increases. 

\section{Results} \label{sec:results}
\subsection{Results Obtained Using the Unbalanced Data Set} \label{subsec:res_unbalanced}

The ROC curves of SVM, SVM-KNN, and AdaBoost on an unbalanced dataset are shown in Figures \ref{fig:rocs}(a), \ref{fig:rocs}(b), and \ref{fig:rocs}(c) respectively. The accuracies of these methods are 98.6\%, 98.86\%, and 97.2\% respectively, shown in Table \ref{table:res_unbalanced}. Notably, the difference between the sensitivity and specificity of SVM and SVM-KNN is approximately 9\%.




\begin{table}[h]
\centering
\caption{Results of classification of unbalanced dataset: F-score is an essential measure of the performance of any classifier applied to an unbalanced dataset, which has been ignored in the available literature.}
\begin{tabular}{|c|c|c|c|c|} \hline
Methods & Accuracy (\%) & Sensitivity & Specificity & F\-score \\ \hline
SVM   &  98.6 &  0.9150  &  0.9937 & 0.9551\\ \hline
SVM-KNN & 98.86 & 0.9159 & 1 & 0.9159 \\ \hline 
AdaBoost  &  97.2  &  0.9012  &  0.9129 & 0.9406 \\ \hline
Asymmetric AdaBoost (new contribution) & 99.99 & 1  &  1  &  1 \\ \hline
\end{tabular}
\label{table:res_unbalanced}
\end{table}
\subsection{Results After Artificially Balancing the Data Set}\label{subsec:res_artbalanced}
\begin{figure}
\begin{center}
\begin{tabular}{cc}

\includegraphics[width=5cm,keepaspectratio]{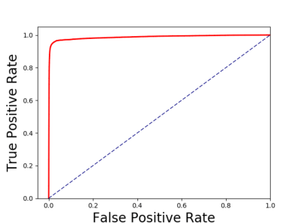} \label{hi} &   \includegraphics[width=5cm,keepaspectratio]{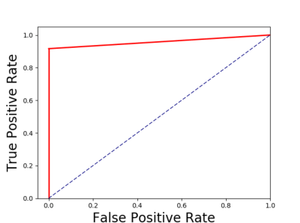} \\    (a) SVM: unbalanced. & (b) SVM-KNN: unbalanced.  \\   AUC = 98.72\%. &AUC = 95.83\%. \\[6pt]
  
\includegraphics[width=5cm,keepaspectratio]{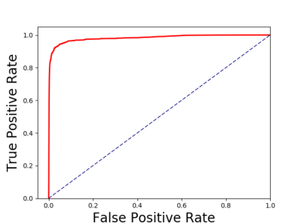} &   \includegraphics[width=5cm,keepaspectratio]{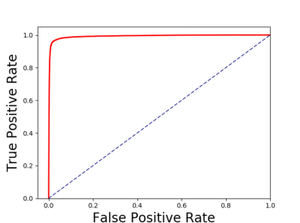} \\    (c) AdaBoost: unbalanced. & (d) SVM: artificially balanced.\\  AUC = 98.23\%. & AUC = 99.27\%. \\[6pt]

\includegraphics[width=5cm,keepaspectratio]{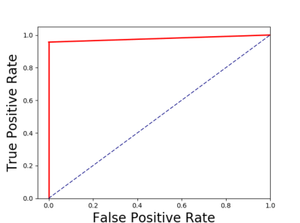} &   \includegraphics[width=5cm,keepaspectratio]{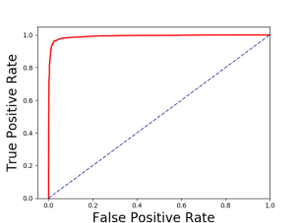} \\    (e) SVM-KNN: artificially balanced. & (f) AdaBoost: artificially balanced.\\ AUC = 97.93\%.& AUC = 99.21\%. \\[6pt]

\multicolumn{2}{c}{\includegraphics[width=5cm,keepaspectratio]{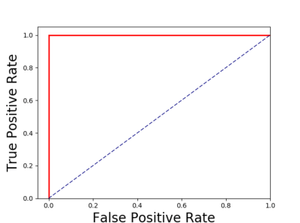}}\\    \multicolumn{2}{c}{(g) Asymmetric AdaBoost: unbalanced. AUC = 100.0\%.}
\end{tabular}
\end{center}
\caption{ROC curves of the different methods explored: all the values for \textit{area under the curve} (AUC) are provided with the plots, for the different cases. Note how the values of AUC are more for the balanced cases, as compared to the unbalanced cases.}
\label{fig:rocs}
\end{figure}

The ROC curves of SVM, SVM-KNN, and AdaBoost after artificial balancing are shown in Figures \ref{fig:rocs}(a), \ref{fig:rocs}(b), and \ref{fig:rocs}(c) respectively. The accuracies of these methods are 96.92\%, 97.87\%, and 96.54\% respectively, shown in Table \ref{table:res_balanced}. Notably, the difference between the sensitivity and specificity of all the models is negligible; in the case of AdaBoost, both the sensitivity and specificity are about 5\% higher compared to the values attained with an unbalanced dataset. In this case, there is no requirement to report the F-Score as it is a metric that should be used in the case of unbalanced or biased datasets. The method of artificial balancing does well to reduce the effects of bias, as seen from the small difference between sensitivity and specificity.
%
%
%
%
%
%
%

\begin{table}[h]
\centering
\caption{Results of classification of balanced dataset: the accuracy of classification drops when data is balanced. }
\begin{tabular}{|c|c|c|c|} \hline
Methods & Accuracy (\%) & Sensitivity & Specificity  \\ \hline
SVM   &  96.92 &  0.9576  &  0.9808 \\ \hline
SVM-KNN & 97.87 & 0.9575 & 1  \\ \hline
AdaBoost  &  96.54  & 0.9663 & 0.9645  \\ \hline
\end{tabular}
\label{table:res_balanced}
\end{table}
\subsection{Results of the Asymmetric AdaBoost Classifier}
The entire dataset was split into training and testing sets. Weights were assigned to both the classes: the stars class was assigned a weight of 0.10 and the weight of the quasar class is kept constant, and equal to 1 (these numbers were selected based on iterative experimentation with different values of initial weights). The mean accuracy of classification was 99.9995\% after running the asymmetric AdaBoost classifier for 1000 iterations. The ROC curve of this method is shown in Figure \ref{fig:rocs}(g).

Simply put, an appropriate weight initialization arrives at the best weight distribution for a given number of estimators faster than equal initial weights. The ROC curve plotted for asymmetric AdaBoost is shown in Figure \ref{fig:rocs}(g). Asymmetric AdaBoost tends to classify positives samples more carefully when compared to negative samples as it corrects the misclassification. Its precision and recall values are found to be equal to 1. The value of F-score is also equal to 1, shown in Table \ref{table:res_unbalanced}.

In the design of any experiment, there exists an inherent trade-off between good results and time of execution. Using asymmetric AdaBoost improves the time of execution, while best-preserving accuracy.
\begin{table}[h]
\centering
\caption{Comparison of accuracies of classification achieved by Gao et al. (2008), Elting et al. (2008), Peng et al. (2013) before and after artificial balancing: accuracy drops after balancing.}
\begin{tabular}{|c|c|c|c|} \hline
Methods & Accuracy before Balancing(\%) & Accuracy after Balancing(\%)  \\ \hline
Gao et al. (2008)   &  97.55 &  96.92  \\ \hline
Elting et al. (2008) &  98.5  &  96.92 \\ \hline
Peng et al. (2013) &  98.85  &  97.87 \\ \hline
\end{tabular}
\label{table:res_comparison}
\end{table}

\section{Discussion}\label{discussion}
We implemented the methods for quasar-star classification which are already reported in the literature literature, with and without artificial balancing. An accuracy of 98.6\% was obtained for SVM and 98.86\% accuracy for the SVM-KNN method without artificial balancing. The artificial balancing of the dataset was accomplished by considering an equal number of quasar and star samples for classification (which is equal to the number of quasars in the dataset, as the quasars' class has lesser number of samples). The accuracy of classification of artificially balanced data drops from 98.6\% to 95.8\% for SVM with linear kernel and 98.86\% to 97.05\% for the SVM-KNN method. This is shown in Tables \ref{table:res_unbalanced} and \ref{table:res_balanced}. 

The choice of classifiers has further been verified by exploring the separability of the data. The data is not separable across the axes of $u-g$, $g-r$, $r-i$, and $i-z$: using any of these four features alone, it is very difficult to discern between the two classes, as the majority of the data points are overlapping, or very close to each other in the feature space, along these axes. However, when considering the redshift ($Z$), we can observe that the data is considerably separable based on this feature. There's a slight overlap, near the edge or corner points of the quasars' class (this can be observed by inspecting the corner points of the convex hull of the quasars' class): since the overlap is very little, SVM is an appropriate method to be explored as a classifier. However, the slight overlap results in accuracy of 96.92\% by SVM (Table\ref{table:res_balanced})and not 100\%. On the other hand, tree-based classifiers work by \textit{multiple recursive partitioning} of the feature space, and hence, in general, are the choice of classifiers for datasets which are mostly linearly inseparable. Since the overlap is not much, with the appropriate initial weights and with the cumulative effect of the remaining features, Asymmetric AdaBoost resulted in an accuracy which is near perfect (Table\ref{table:res_unbalanced})!
\section{Conclusion}
Asymmetric AdaBoost is endowed with greater computational efficacy compared to SVM. Given high accuracy, fast speed and easy modulation of parameters in contrast to SVM, asymmetric AdaBoost is a good choice as a classifier as specified in Tables \ref{table:res_unbalanced} and \ref{table:res_balanced}.

The approaches explored in this paper can be used to solve the star-quasar classification problem in particular, and other problems in astronomy in general. These classifiers can be used to classify multi-wavelength astronomical data sources and pre-select quasar candidates for large surveys. The paper is firmly focused on scientific correctness and algorithmic relevance. Different ML approaches have been discussed and should be interpreted in that light, not as a suite of trial and error approaches to pick the better ones.


\begin{thebibliography}{99}

\bibitem{Abazajian:c2} Abazajian K. N. et al., The Seventh Data Release of the Sloan Digital Sky Survey, Astrophysical Journal Supplement 182 (2009) 543-558, doi:10.1088/0067-0049/182/2/543

\bibitem{Adelman:c1} 
Adelman-McCarthy J. K., et al., The Sixth Data Release of the Sloan Digital Sky Survey, Astrophysical Journal Supplement 175 (2008) 297-313, doi:10.1088/0067-0049/182/2/543


\bibitem{Basak :c11}Basak S, Saha S et al., 2016, Star Galaxy Separation using Adaboost and Asymmetric Adaboost, doi: 10.13140/RG.2.2.20538.59842

 


\bibitem{Elting:c6} Elting C., Bailer-Jones C. A. L., Smith K. W., 2008, Photometric Classification of Stars, Galaxies and Quasars in the Sloan Digital Sky Survey DR6 Using Support Vector Machines, AIP Conference Proceedings, Volume 1082, Issue 1, doi:10.1063/1.3059095

\bibitem{Freund:c8} Freund Y.,  Schapire R. E., 1996, Experiments with a New Boosting algorithm, Proceedings of 13th International Conference on Machine Learning (ICML), Bari, 3-6 July 1996, 148-156

\bibitem{Gao:c5} Gao D., Zhang Y., Zhao Y, 2008, Support vector machines and kd-tree for separating quasars from large survey data bases, MNRAS, 386: 1417-1425, doi:10.1111/j.1365-2966.2008.13070.x

\bibitem{Hassanat et al.} Hassanat A.~B., Abbadi M. A., Altarawaneh G.~A., Alhasnat A.~A., 2014, Solving the Problem of the K Parameter in the KNN Classifier Using an Ensemble Learning Approach, preprint(arxiv:1409.0919v1)

\bibitem{Alba-Castro(2012):c9} Landesa-V{\'{a}}zquez I, Alba-Castro J. L., 2012, Shedding light on the asymmetric learning capability of AdaBoost, Pattern Recognition Letters 33(3), p.247-255, doi:10.1016/j.patrec.2011.10.022


\bibitem{Peng et al.(2013):c7} Peng N., Zhang Y., Zhao, Y., 2013, A SVM-kNN method for quasar-star classification, Sci. China Phys. Mech. Astron. 56: 1227, doi:10.1007/s11433-013-5083-8

\bibitem{sdss}http://skyserver.sdss.org/casjobs

\bibitem{cosmos} N.C.Hambly,  M.J.  Irwin,   H.T. MacGillivray, 2001, The SuperCOSMOS Sky Survey – II. Image detection, parametrization, classification and photometry, Monthly Notices of the Royal Astronomical Society. 326. 1295-1314. 10.1111/j.1365-2966.2001.04661.x.

\bibitem{Eduardo et al. (2017):c13}A. A. Miller and M. K. Kulkarni and Y. Cao and R. R. Laher and F. J. Masci and J. A. Surace, 2017, Preparing for Advanced LIGO: A Star–Galaxy Separation Catalog for the Palomar Transient Factory, The Astronomical Journal, Volume 153, Number 2, Pages 73



\end{thebibliography}
\end{document}